\documentclass{jkas}

\def\year{2015}
\def\month{Dec} 
\def\volume{00} 
\def\issue{0} 
\def\beginpage{1} 
\def\endpage{99} 
\def\received{December 1, 2015} 
\def\accepted{December 15, 2015} 
\date{Received \received; accepted \accepted}

\newcommand{\kms}{\rm km~s^{-1}}

\usepackage{color}
\usepackage{multirow}
\usepackage{amsmath}
\usepackage{afterpage}

\title{
COMPACT GROUPS OF GALAXIES WITH COMPLETE SPECTROSCOPIC REDSHIFTS IN THE LOCAL UNIVERSE
}

\author[1,3]{Jubee Sohn}
\author[2]{Ho Seong Hwang}
\author[3]{Margaret J. Geller}
\author[4,5]{Antonaldo Diaferio}
\author[6]{Kenneth J. Rines}
\author[1]{Myung Gyoon Lee}
\author[1]{Gwang-Ho Lee}

\affil[1]{Department of Physics and Astronomy, Seoul National University, Gwanak-gu, Seoul 88226, Korea; \email{jsohn@cfa.harvard.edu}}
\affil[2]{School of Physics, Korea Institute for Advanced Study, 85 Hoegiro, Dongdaemun-Gu, Seoul 02455, Korea}
\affil[3]{Smithsonian Astrophysical Observatory, 60 Garden Street, Cambridge, MA 02138, USA}
\affil[4]{Dipartimento di Fisica, Universit\`a degli Studi di Torino, V. Pietro Giuria 1, I-10125 Torino, Italy}
\affil[5]{ Istituto Nazionale di Fisica Nucleare (INFN), Sezione di Torino, V. Pietro Giuria 1, I-10125 Torino, Italy}
\affil[6]{Department of Physics and Astronomy, Western Washington University, Bellingham, WA 98225, USA} 

\def\corrauthor{
J. Sohn
}

\def\runningauthor{
Sohn et al.
}

\def\runningtitle{
A Spectroscopically Confirmed Compact Group Catalog
}

\def\keywords{
galaxies:evolution -- galaxies: groups: general -- galaxies: interactions -- galaxies
}

\def\abstracttext{
Dynamical analysis of compact groups provides 
 important tests of models of compact group formation and evolution. 
By compiling 2066 redshifts from FLWO/FAST, 
 from the literature, and from SDSS DR12
 in the fields of compact groups in \citet{McC09}, 
 we construct the largest sample of compact groups 
 with complete spectroscopic redshifts in the redshift range $0.01 < z < 0.22$.
This large redshift sample shows that 
 the interloper fraction in the \citet{McC09} compact group candidates is $\sim 42\%$.
A secure sample of 332 compact groups
 includes 192 groups with four or more member galaxies and 
 140 groups with three members.
The fraction of early-type galaxies in these compact groups is 62\%,
 slightly higher than for the original Hickson compact groups. 
The velocity dispersions of early- and late-type galaxies in compact groups 
 change little with groupcentric radius; the radii sampled are less than $100 ~h^{-1}$ kpc, 
 smaller than the radii typically sampled by members of massive clusters of galaxies.
The physical properties of our sample compact groups 
 include size, number density, velocity dispersion, and local environment; 
 these properties slightly differ from those derived 
 for the original Hickson compact groups and for the DPOSS II compact groups. 
Differences result from subtle differences in the way 
 the group candidates were originally selected.
The space density of the compact groups 
 changes little with redshift over the range covered by this sample. 
The approximate constancy of the space density for this sample is
 a potential constraint on the evolution of compact groups on a few Gigayear timescale.
}

\begin{document}
\jkashead 


\section{INTRODUCTION\label{sec:intro}}

Compact groups of galaxies provide
 a very dense environment for the study of galaxy evolution. 
These groups contain 
 a few galaxies 
 separated by projected distances of only a few tens of kiloparsec,
 comparable with the galaxy sizes.
Compact groups are thus the densest galaxy systems known.
The line-of-sight velocity dispersions of these groups
 ($\sim 200~ \kms$, \citealp{Hic92})
 are lower than those of clusters 
 ($500 - 1000~ \kms$, \citealp{Rin06,Hwa12}),
 but comparable with many loose groups \citep{Ein03}.  
The high density, low velocity dispersion, and short crossing time of compact groups
 make them a test-bed for the study of  galaxy interactions
 (e.g. \citealp{Hic92, MdO94, Bit11, Soh13, Bit14, Fed15}).

The physical processes important for
 the formation and evolution of compact groups remain unclear.
The mere survival of these systems for times
 much longer than a few crossing times has been a long-standing puzzle.
Several numerical simulations showed that
 galaxies within a compact group should merge and
 the group should thus disappear \citep{Bar85,Bar89,Mam87}.
In fact, \citet{Bar89} proposed that 
 compact group galaxies merge into a single elliptical galaxy
 on a very short time scale ($< 0.02$ Hubble time),
 comparable with the observed crossing time \citep{Hic92,Pom12}.
Other simulations suggested that
 compact groups can survive much longer than the crossing time
 \citep{Gov91, Ath97}.
\citet{Gov91} showed that
 for galaxies with a mass range appropriate to compact groups,
 some group members may remain in quasi-stable orbits for billions of years.
\citet{Ath97} suggested that
 compact groups survive because the galaxies are embedded in a common halo. 
In yet another picture,
 \citet{Dia94} proposed that
 compact groups form within a single rich loose group and
 they can thus acquire new members from the surrounding environment
 thus lengthening their apparent lifetime.
The number density of compact groups as a function of epoch
 may thus depend not only on the merger rate of galaxies within them
 but also on the replenishment with new members accreted from the surroundings.

The environments of compact groups are
 an important clue to understand their formation and evolution.
Known compact groups inhabit a range of environments
 ranging from clusters and rich groups to low density regions.
\citet{Ram94} found that
 76\% of a sample of 38 Hickson compact groups are embedded in rich groups. 
Several studies showed that 
 a significant fraction of compact groups 
 are embedded in clusters, rich groups, less dense poor groups and
 in the surrounding larger-scale structures \citep{Roo94, Rib98, And05, Men11, Pom12}. 
In some of these studies neither the compact group candidates nor the environments
 have complete redshift measurements (e.g. see the discussion by \citealp{Men11}). 
A fuller understanding of the environmental issues
 affecting the formation and evolution of compact groups
 requires complete spectroscopy of compact group candidates
 within a large volume redshift survey. 
 
The abundance of compact groups as a function of redshift is
 also a potential constraint on the evolution of these systems.
For example,
 \citet{Kro15} suggested that 
 the abundance of compact groups should decline significantly over a 1 Gyr timescale
 for halos composed of exotic dark matter particles. 
The suggestion by \citet{Kro15} tacitly assumes
 that compact groups do not accrete new members from the environment
 in contrast with the model proposed by \citet{Dia94}. 
To date there are no direct observational measures
 of the space density evolution of compact groups to test these conjectures.  
 
There have been several attempts to construct larger catalogs of compact groups
 \citep{Ros77,Hic82,Pra94,Iov03,Lee04,deC05,McC09}. 
\citet{Hic82} published a widely used catalog of 100 compact groups.
\citet{McC09} used Hickson's criteria to identify compact groups 
 in the photometric data of the Sloan Digital Sky Survey (SDSS) data release 6 
 (DR6, \citealp{Ade08}). 
Currently the sample of \citet{McC09} is the largest catalog of compact group candidates
 with 77,088 tentative groups.
However, at least 55\% of the compact group candidates
 from the magnitude-limited sample of $14.5 \leq r \leq 18.0$
 could be contaminated by interlopers
 as a result of their selection based on photometric data.
This interloper fraction may be greater
  for their faint sample compact groups with $14.5 \leq r \leq 21.0$. 

Redshift surveys of compact group candidates provide
 a basis for cleaner catalogs
 better suited to testing models 
 for the formation and evolution of these systems \citep{Hic92,Pom12}. 
\citet{Hic92} showed that
 69 of the 100 compact groups in his original catalog 
 include four or more members with accordant redshifts. 
Similarly, 
 \citet{Pom12} observed 138 compact group candidates
 drawn from the second digital Palomar Observatory Sky Survey \citep{Iov03,deC05}; 
 96 of these contain three or more galaxies with accordant redshifts 
 (DPOSS II compact groups hereafter).
The 70\% success rate for these two catalogs underscores 
 the importance of spectroscopic observations 
 for constructing a robust sample of compact groups. 
 
We conduct a spectroscopic survey of compact group candidates in the SDSS DR6
 to construct an updated sample of compact groups with complete redshifts.
By adding 2066 redshifts, 
 we construct the largest catalog of compact groups with complete spectroscopic redshifts.
Based on this sample, 
 we examine the physical properties of compact groups including
 size, velocity dispersion, space density and local environment. 
We compare these properties with the Hickson and the DPOSS II groups.  
We show that the physical characteristics
 of the groups in our catalog are not a strong function of the redshift of the system. 
We also estimate the space density of compact groups
 as a function of redshift for the range $0.01 < z < 0.22$.
These estimates are a first step toward using the space density
 as a test of models for the evolution of these systems.

Section 2 describes the basic sample used to construct the compact group catalog. 
Section 3 explains the method of identifying compact group members
 once the redshifts are measured.
We examine the physical properties of the sample compact groups with complete redshifts 
 and compare them with other compact group catalogs in \S 4.
We summarize in \S 5. 
Throughout, we adopt  $\Lambda$CDM cosmological parameters 
 $H_{0} = 100~ h ~\rm km ~\rm s^{-1} ~\rm Mpc^{-1}$, 
 $\Omega_{\Lambda}=0.7$, and
 $\Omega_{m}=0.3$. 
We use $H_{0} = 100~ h ~\rm km ~\rm s^{-1} ~\rm Mpc^{-1}$
 to facilitate comparison with earlier work.

\section{DATA}

\subsection{Parent Sample}
\citet{McC09} used Hickson's criteria \citep{Hic82} 
 to identify compact group candidates 
 in the photometric sample of SDSS DR6 galaxies. 
Hickson's criteria can be expressed as follows:
 $N (\Delta m < 3) \geq 4$, $R_{N} \geq 3 R_{G}$, and 
 $\mu_{gr} < 26.0~ {\rm mag~arcsec}^{-2}$. 
$N (\Delta m < 3)$ means 
 the total number of member candidate galaxies 
 within 3 mag of the brightest galaxy in a system.
$R_{G}$ is the angular size of the smallest circle containing all members, 
 and $R_{N}$ is the angular size of the largest circle that includes no additional galaxies 
 within 3 mag of the brightest galaxy. 
$\mu_{gr}$ is the mean surface brightness within the circle of radius $R_{G}$. 
\citet{McC09} constructed two catalogs of compact group candidates
 with different magnitude limits:
 catalog A with $14.5 \leq r \leq 18.0$ and 
 catalog B with $14.5 \leq r \leq 21.0$. 
Catalog A and catalog B list 2297 and 74,791 compact group candidates
 with 9713 and 313,508 tentative member galaxies, respectively. 

Here we use catalog A as a parent sample.
Despite the bright magnitude limit for catalog A,
 only a small fraction of group candidates in the catalog
 were previously confirmed as genuine compact groups
 based on spectroscopic redshifts. 
There are only 70 complete compact groups in the catalog 
 containing four or more members 
 with velocity difference from the mean group velocity
 less than $1000~\kms$ (see Table \ref{tab:obsstat}).

\subsection{Redshift Data}

\begin{figure}
\centering
\includegraphics[width=80mm]{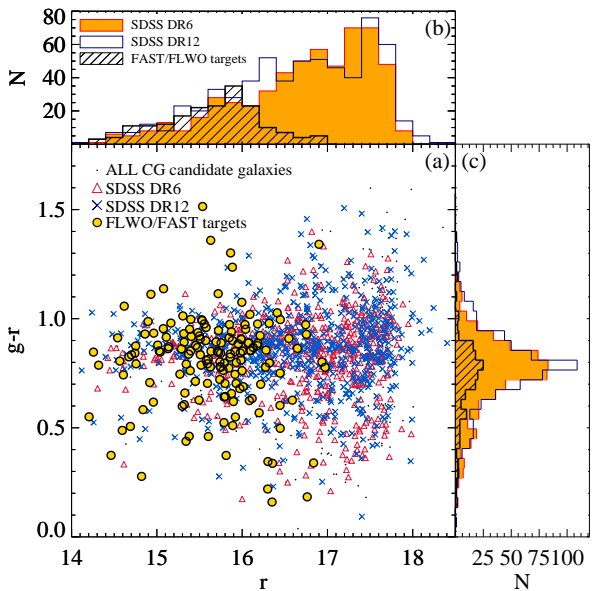}
\caption{
(a) The $g-r$ vs. $r$ color magnitude diagram of FLWO/FAST target galaxies 
 compared with
 compact group galaxies from SDSS DR6 (triangles, \citealp{McC09}) and SDSS DR12 (crosses). 
 Dots indicate compact group candidate galaxies without redshifts.  
(b)-(c) The $r$-band magnitude and $g-r$ color distributions for 
 compact group galaxies with SDSS DR6 redshifts (filled histogram), 
 those with SDSS DR12 redshifts (open histogram), and 
 FLWO/FAST target galaxies (hatched histogram). }
\label{cmd}
\end{figure}

To construct a sample of compact groups with complete redshifts, 
 we conducted a redshift survey of the galaxies 
 in the fields of compact group candidates 
 in catalog A of \citet{McC09} (see their Table 3). 
Among the candidate group galaxies, 
 we primarily targeted galaxies in groups 
 that already have two or three members with measured redshifts. 
We then ranked the targets by their apparent magnitude; 
 the targets have $r-$band magnitudes in the range $14.2 < r < 17.0$.
To avoid other selection effects, we did not use any selection criteria other than apparent magnitude. 
Figure \ref{cmd} shows the color-magnitude diagram for the target galaxies.  
We use the extinction-corrected Petrosian magnitudes from the SDSS DR12. 
We also plot the compact group galaxies 
 identified with  SDSS DR6 and DR12 redshift data. 
The FAST target galaxies are generally brighter than
 compact group galaxies with the SDSS redshifts. 
The color distribution of compact group galaxies  peaks at $g-r \sim 0.8$. 

We obtained long-slit spectra of 193 galaxies 
 with the FAST spectrograph \citep{Fab98} 
 installed on the 1.5m Tillinghast telescope
 at the Fred Lawrence Whipple Observatory (FLWO)
 from 2013 May to 2014 May.
We used a long slit with a 3 arcsec width and a 300 line grating 
 providing spectral resolution of $2.94~{\rm \AA}$ and 
 a dispersion of $1.47~\rm \AA$ pixel$^{-1}$. 
The spectra cover the wavelength range $3470-7420~\rm\AA$. 
The exposure times range from 900 to 1800 s 
 depending on the brightness of the target galaxy. 
We reduced the data using IRAF. 
We measure the redshift of each galaxy with
 the {\it rvsao} \citep{Kur98} task. 
During the pipeline processing, 
 we assign a quality flag of `Q' for high-quality redshifts,
 '?' for marginal cases, and 'X' for poor fits.  
We obtained 193 spectra in this study. 
Among these,
 five have an 'X' flag and three have a '?' flag.
We exclude these eight objects from the analysis. 
The typical velocity measurement error for 
 the FAST spectra of compact group galaxies is $22~\kms$.
 
We supplement these data with redshifts 
 from the literature (see \citealp{Hwa10} for details) 
 including the FAST archive and the SDSS DR12 \citep{Ala15}.
There are two redshifts from FAST observations between 2006 and 2008 measured as part of 
 an unpublished study of low-redshift clusters and groups (P.I.: K. Rines).
There are 161 and 1718 new redshifts from the literature and SDSS DR12
 for galaxies in the fields of compact group candidates in \citet{McC09}. 
The total number of redshifts we add to the \citet{McC09} catalog is 2066.

\section{A SAMPLE OF COMPACT GROUPS WITH REDSHIFTS}

We combine the 2066 additional redshifts 
 with the existing data for galaxies in the fields of compact group candidates
 in \citet{McC09}.
We determine compact group membership based only on galaxies 
 with a spectroscopic redshift.

\begin{table*}[t!]
\caption{Statistics of Sample Compact Groups\label{tab:obsstat}}
\centering
\begin{tabular}{lcccc}
\toprule
Survey & $N\geq4$ CGs & $N=3$ CGs & $N\geq3$ incomplete CGs \\
           & (members) & (members) & (members)  \\
\midrule
\citet{McC09} &  70 (291) &  55 (165) & 191 (573)  \\
SDSS DR12 	  & 164 (685) & 125 (375) & 346 (1038)  \\
This study    & 192 (799) & 140 (420) & 395 (1185)  \\
\bottomrule
\end{tabular}
\end{table*}

We first compute the median redshift of compact group member candidates 
 as a tentative group redshift. 
We then calculate the line-of-sight velocity differences 
 between member candidates and the median redshift, and
 remove foreground and background galaxies 
 with line-of-sight velocity differences larger than $1500 ~\kms$. 
We use the mean velocity of the remaining group galaxies 
 as a group systemic redshift. 
We finally select member galaxies in each group
 with concordant redshifts of $|v_{galaxy} - v_{group}| \leq 1000 ~\kms$, 
 following the velocity separation of CG galaxies 
 used in previous studies \citep{Hic92, Men11, Pom12}.  
To check the reliability of the cutoff velocity $1000~\kms$, 
 we test the group selection 
 with the larger cutoff velocities of $1500~\kms$ and $2000~\kms$. 
With the larger cutoff velocities, 
 only a few additional compact group candidates are newly identified as true compact groups. 
Thus we use cutoff velocity $1000~\kms$ 
 for direct comparison with other compact groups based on the same cutoff. 

We adopt galaxy morphology information for the compact group galaxies 
 from the Korea Institute for Advanced Study (KIAS) DR7 value-added catalog (VAGC) \citep{Choi10}. 
\citet{Choi10} classified early- and late-type galaxies 
 using the $u-r$ color, the $g-i$ color gradient, 
 and the $i$ band concentration index 
 following the automatic classification scheme suggested by \citet{Par05}. 
We visually classify the morphology of galaxies 
 not included in the KIAS DR7 VAGC using the SDSS images. 

There are three types of compact groups
 in our spectroscopic sample:
 groups with four or more members ($N\ge4$ compact groups hereafter), 
 groups with three members ($N=3$ compact groups hereafter), and
 incomplete groups with three confirmed members plus 
 one or more tentative member galaxies 
 with unknown redshifts ($N\geq3$ incomplete compact groups hereafter).
We do not use these $N\geq3$ incomplete groups for further analysis 
 except when computing the group abundance.
It is unclear whether 
 these $N\geq3$ incomplete groups would be confirmed as $N\geq4$ or $N=3$ groups; 
 thus we do not include them in the analysis. 
However, these incomplete groups remain useful 
 for determining the abundance of $N\ge3$ compact groups. 
The incomplete compact groups consist of at least three member galaxies and 
 this satisfy our compact group selection criteria. 
Thus, we include them only when we compute the group abundance. 
 
\citet{Hic82} originally defined compact groups 
 with $N\ge4$ members rather than with $N\ge3$ members. 
\citet{Dup13} compared the properties (i.e. stellar mass, star formation rate and color) 
 of compact triplets with those of larger compact group candidates. 
They concluded that galaxy triplets do not differ from more populated compact groups,
 but they do differ from galaxy pairs and clusters. 
Many previous studies have included $N=3$ compact groups 
 when all three galaxies have measured redshifts confirming their membership.
We thus include the $N=3$ compact groups.

Table \ref{tab:obsstat} summarizes our compact group selection. 
In the original catalog of \citet{McC09}, 
 there are 70 $N\ge4$ and 55 $N=3$ genuine compact groups 
 with 291 and 165 members, respectively.
By adding the SDSS DR12 data, 
 the number of compact groups increases to 
 164 $N\ge4$ and 125 $N=3$ compact groups
 with 685 and 375 members, respectively. 
Finally our FLWO/FAST observations contribute an
 additional 28 $N\ge4$ and 15 $N=3$ complete compact groups 
 for a final sample of 
 192 $N\ge4$ and 140 $N=3$ compact groups with 
 799 and 420 member galaxies, respectively.
Among the $N\ge4$ compact groups, 
 there are 164, 26, 1 and 1 groups with $N=4,5,6$ and 7 members.
We also identified $395~N\geq3$ incomplete compact groups 
 with a total of 1185 member galaxies. 
The number of compact groups with complete spectroscopic redshifts in this study
 is about three times larger than the number in the original \citet{McC09} catalog.

The new redshift data also identify 
 many chance alignments among the compact group candidates of \citet{McC09}.
There are 144 and 9 compact group candidates 
 that turn out to be chance alignments of galaxies with discordant redshifts
 based on the SDSS DR12 and FLWO/FAST data, respectively. 
This substantial number of chance alignments clearly underscores
 the importance of spectroscopic redshifts 
 for reducing the contamination the compact group sample. 
 
\begin{table}[t!]
\caption{Interloper Statistics for Compact Groups\label{tab:inter}}
\centering
\begin{tabular}{lccc}
\toprule
Redshift Survey & $N_{tot}^{\rm a}$ & $N_{int}$ & $f_{int}^{\rm a}$ \\
\midrule
\citet{McC09} &  958  & 503 & $52.5 \pm 1.6\%$  \\
SDSS DR12     & 1883 & 815 & $43.3 \pm 1.1\%$  \\
FLWO/FAST    &  242  &   83 & $34.3 \pm 3.2\%$  \\
Total              & 2125 & 898 & $42.3 \pm 1.1\%$  \\
\bottomrule
\end{tabular}
\tabnote{$^{\rm a}$ The number of galaxies in $N\ge4$ and $N=3$ compact groups, and chance alignments. }
\tabnote{$^{\rm b}$ Error in the interloper fraction is 1$\sigma$ standard deviation derived from 1000 bootstrap resamplings.}
\end{table}
 
The data yield a measure of the interloper fraction for our sample groups.
These interlopers are galaxies initially selected as candidate group members,
 but the redshift shows that they are non-members.
We define the interloper fraction as 
\begin{equation}
 f_{int} = 1 - \frac {N_{\rm members}} {N_{\rm total~candidates}},
\end{equation} 
 where $N_{\rm members}$ is the number of spectroscopically confirmed members and 
  $N_{\rm total~candidates}$ is the number of compact group candidate galaxies
  in the fields of $N\ge4$ compact groups, $N=3$ compact groups, and chance alignments. 
We base the estimate of the interloper fraction
 on the $N\ge4$ and $N=3$ compact groups and chance alignments
 where most candidate group galaxies in \citet{McC09} have measured redshifts.
We exclude $N\geq3$ incomplete compact groups and group candidates,
 because we do not know the exact number of interlopers in these groups.
Table \ref{tab:inter} lists the numbers of compact group candidate galaxies and 
 the number of interlopers.
The interloper fraction we estimate for 
 the original \citet{McC09} catalog is $52.5 \pm 1.6\%$,
 consistent with their estimate of 55\%.
The error in the interloper fraction is the $1\sigma$ standard deviation
 in the interloper fraction obtained with 1000 bootstrap resamplings.
The interloper fractions for the groups we complete with the SDSS DR12 and FLWO/FAST data 
 are slightly smaller than the estimate for the \citet{McC09} catalog.
We add many redshifts of bright galaxies from FLWO/FAST and SDSS DR12 data sets 
 that are more likely to be true members of the groups than
 the fainter candidates also included in the \citet{McC09} estimate (see Figure \ref{cmd}). 
 
\afterpage{
\landscape
\begin{table}[t!]
\caption{A Catalog of Spectroscopically Identified 
 Compact Groups$^{\rm a}$\label{tab:grcat}}
\centering
\begin{tabular}{lrrccccccr}
\toprule
\multirow{2}{*}{ID$^{\rm b}$} & R.A. & Decl. &
\multirow{2}{*}{$n_{mem}$} & \multirow{2}{*}{z$^{\rm c}$} & $R_{gr}^{\rm c}$ & $R_{gr}^{\rm c}$ & $\log \rho^{\rm c}$ & $\sigma^{\rm c}$ & \multirow{2}{*}{Nearby cluster$^{\rm d}$} \\
	& (J2000) & (J2000) & 	& 		  & (arcmin) & ($h^{-1}$ kpc) & ($h^{3}$ Mpc$^{-3}$) & (km s$^{-1}$) \\
\midrule
SDSSCGA00027 &   5.91056 &  -0.78910 & 3 & $0.0633 \pm 0.0003$ & $0.325 \pm 0.070$ & $16.6 \pm 3.6$ & $5.19 \pm 4.60$ & $140 \pm  20$ &                    SDSS-C42022 \\
SDSSCGA00029 & 204.18318 &  -3.49931 & 3 & $0.0531 \pm 0.0000$ & $0.118 \pm 0.027$ & $ 5.1 \pm 1.2$ & $6.72 \pm 6.17$ & $ 14 \pm   6$ &                            \\
SDSSCGA00035 & 141.03156 &  13.21444 & 4 & $0.0780 \pm 0.0009$ & $0.433 \pm 0.076$ & $26.9 \pm 4.7$ & $4.69 \pm 4.14$ & $500 \pm  69$ &                            \\
SDSSCGA00037 &  10.36639 &  -9.23039 & 3 & $0.0470 \pm 0.0001$ & $0.263 \pm 0.077$ & $10.2 \pm 3.0$ & $5.83 \pm 5.28$ & $ 76 \pm  25$ &                            \\
SDSSCGA00042 & 155.54219 &  38.52117 & 4 & $0.0549 \pm 0.0007$ & $0.586 \pm 0.131$ & $26.3 \pm 5.9$ & $4.72 \pm 4.18$ & $470 \pm  64$ &                 400dJ1020+3831 \\
SDSSCGA00046 & 127.02769 &  44.76412 & 4 & $0.1465 \pm 0.0004$ & $0.267 \pm 0.076$ & $28.7 \pm 8.2$ & $4.61 \pm 4.11$ & $320 \pm  73$ &                      Abell0667 \\
SDSSCGA00070 & 157.91676 &  36.01777 & 4 & $0.0861 \pm 0.0002$ & $0.401 \pm 0.073$ & $27.2 \pm 5.0$ & $4.68 \pm 4.14$ & $180 \pm  29$ &             NSCSJ103122+355649 \\
SDSSCGA00071 &  31.82000 &  -1.01116 & 4 & $0.1181 \pm 0.0010$ & $0.383 \pm 0.072$ & $34.3 \pm 6.5$ & $4.37 \pm 3.81$ & $600 \pm 130$ &                            \\
SDSSCGA00090 & 141.44824 &   7.72078 & 4 & $0.1360 \pm 0.0008$ & $0.333 \pm 0.075$ & $33.6 \pm 7.6$ & $4.40 \pm 3.90$ & $540 \pm 120$ &                            \\
SDSSCGA00110 & 147.18958 &  25.49768 & 3 & $0.0455 \pm 0.0000$ & $0.344 \pm 0.072$ & $12.9 \pm 2.7$ & $5.52 \pm 4.95$ & $ 22 \pm   7$ &                            \\
\bottomrule
\end{tabular}
\tabnote{
$^{\rm a}$ 
The complete table is available on-line at \url{http://astro.snu.ac.kr/~jbsohn/compactgroups/}.
A portion is shown here for guidance regarding its form and content.
\\$^{\rm b}$ ID from Table 1 in \citet{McC09}.
\\$^{\rm c}$ Errors represent the 1-$\sigma$ standard deviation obtained from by resampling the galaxy sample 1000 times. 
\\$^{\rm d}$ Known galaxy clusters in NED at $R_{projected}< 1 h^{-1}$ Mpc from the group center.}
\end{table}
\endlandscape}

Table \ref{tab:grcat} lists 332 compact groups with $N\ge3$ 
 including ID, R.A., Decl., number of members, group redshift,
 group size, group number density, velocity dispersion, 
 and neighboring clusters if there are any. 
The group center is the 
 mean R.A., Decl. and redshift of the member galaxies.
We examine whether 
 there are any compact groups close to galaxy clusters
 using the NASA Extragalactic Database (NED) with the criteria
 $|v_{group} - v_{cluster}| < 3000~\kms$ and $R_{projected} < 1 h^{-1}$ Mpc, 
 typical virial radius ($R_{200}$) for galaxy clusters \citep{Rin13}. 
Table \ref{tab:galcat} lists 1473 galaxies in the fields of compact groups
 in Table \ref{tab:grcat}
 including ID, R.A., Decl., morphology, $r-$band magnitude, $g-r$ color, membership flag, 
 redshifts and its source. 
We list only the galaxies originally included in the compact group catalog of \citet{McC09}.

\begin{table*}[t!]
\caption{A Catalog of Galaxies in 
Spectroscopically Confirmed Compact Groups$^{\rm a}$\label{tab:galcat}}
\centering
\begin{tabular}{lrrcccccc}
\toprule
ID$^{\rm b}$ & R.A. (J2000) & Decl. (J2000) &
Morph.$^{\rm c}$ & $r$ & $g-r$ & 
Member$^{\rm d}$ & $z$ & z source \\
\midrule
SDSSCGA00027.1 &   5.90833 &  -0.78417 & 1 & 15.42 & 0.85 & 1 & $0.0636 \pm 0.00002$ & SDSS \\
SDSSCGA00027.2 &   5.91375 &  -0.79242 & 1 & 16.47 & 0.90 & 1 & $0.0636 \pm 0.00002$ & SDSS \\
SDSSCGA00027.3 &   5.90958 &  -0.79072 & 1 & 16.53 & 0.90 & 1 & $0.0627 \pm 0.00001$ & SDSS \\
SDSSCGA00027.4 &   5.91083 &  -0.79033 & 1 & 16.94 & 1.16 & 0 & $0.2724 \pm 0.00007$ & SDSS \\
SDSSCGA00029.1 & 204.18459 &  -3.49792 & 1 & 14.90 & 0.83 & 1 & $0.0531 \pm 0.00015$ &  NED \\
SDSSCGA00029.2 & 204.18333 &  -3.50086 & 2 & 15.59 & 0.79 & 1 & $0.0531 \pm 0.00008$ & FLWO \\
SDSSCGA00029.3 & 204.18167 &  -3.49914 & 1 & 16.12 & 1.08 & 1 & $0.0530 \pm 0.00002$ & SDSS \\
SDSSCGA00029.4 & 204.17500 &  -3.50419 & 1 & 17.58 & 0.86 & 0 & $0.0870 \pm 0.00002$ & SDSS \\
SDSSCGA00035.1 & 141.03000 &  13.21414 & 1 & 15.52 & 0.88 & 1 & $0.0789 \pm 0.00001$ & SDSS \\
SDSSCGA00035.2 & 141.03749 &  13.21878 & 2 & 14.32 & 0.90 & 1 & $0.0764 \pm 0.00012$ & FLWO \\
\bottomrule
\end{tabular}
\tabnote{
$^{\rm a}$ The complete table is available on-line at \url{http://astro.snu.ac.kr/~jbsohn/compactgroups/}.
A portion is shown here for guidance regarding its form and content.
\\$^{\rm b}$ ID from Table 3 in \citet{McC09}.
\\$^{\rm c}$ Morphology flag : 1 for early-type galaxies, 2 for late-type galaxies. 
\\$^{\rm d}$ Membership flag : 1 for members, 0 for non-members, 9 for those without redshifts.}
\end{table*} 

In Table \ref{tab:reject}, 
 we also list 139 galaxies with FLWO/FAST redshifts,
 not included in the $N\ge4$ and $N=3$ compact groups 
 listed in Table \ref{tab:grcat}. 
Among the 139 galaxies, 
 49 are in $N\geq3$ incomplete compact groups and 
 nine galaxies are in chance alignments.
The other 81 galaxies are in compact group candidates 
 that could be confirmed as groups if we secure redshifts for the other member galaxies.
These groups require further spectroscopy. 

\begin{table*}[t!]
\centering
\caption{A Catalog of FLWO/FAST Target Galaxies, 
Non-Members of Compact Groups$^{\rm a}$\label{tab:reject}}
\begin{tabular}{lcccccc}
\toprule
ID$^{\rm b}$ & R.A. (J2000) & Decl. (J2000) & $r$ & $g-r$ & $z$ \\
\midrule
SDSSCGA00012.1 & 116.18042 &  16.92258 & 15.36 & 0.92 & $0.0751 \pm 0.00009$ \\
SDSSCGA00012.2 & 116.17667 &  16.92736 & 15.95 & 0.85 & $0.0717 \pm 0.00008$ \\
SDSSCGA00021.1 & 178.52542 &   3.92100 & 15.32 & 0.86 & $0.0747 \pm 0.00001$ \\
SDSSCGA00023.2 & 130.04375 &   8.99811 & 15.75 & 0.96 & $0.0662 \pm 0.00015$ \\
SDSSCGA00067.1 & 176.31375 &  11.49378 & 15.08 & 1.13 & $0.1141 \pm 0.00002$ \\
\bottomrule
\end{tabular}
\tabnote{
$^{\rm a}$ 
The complete table is available on-line at \url{http://astro.snu.ac.kr/~jbsohn/compactgroups/}.
A portion is shown here for guidance regarding its form and content.
\\$^{\rm b}$ ID from Table 3 in \citet{McC09}.}
\end{table*}

\section{COMPACT GROUP PROPERTIES}

\subsection{The Compact Groups}

\subsubsection{Physical Properties}

\begin{figure}
\centering
\includegraphics[width=85mm]{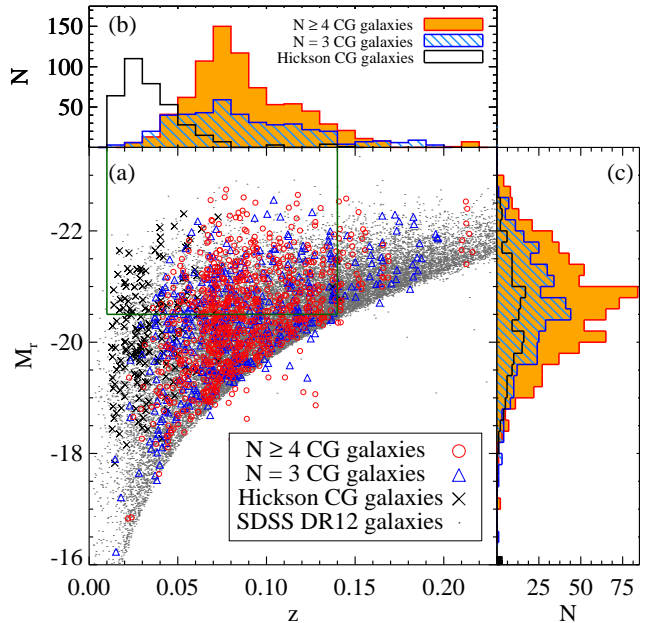}
\caption{
(a) $M_{r}$ -- $z$ diagrams for the galaxies in $N\geq4$ compact groups (circles) and 
 $N=3$ complete compact groups (triangles) in our catalog,
  and for $N\geq3$ Hickson compact group galaxies (crosses). 
Small dots indicate SDSS DR12 galaxies 
 (we display only 1\% of the data for clarity). 
The box defines a volume-limited sample of SDSS DR12 galaxies 
 used for computing surrounding galaxy densities (see Section 4).  
(b) The redshift distributions and (c) the $M_{r}$ distributions for
 $N\geq4$ (filled histogram) and $N=3$ compact groups (hatched histogram) in our sample
 and for the Hickson compact group galaxies (open histogram), respectively.}    
\label{volume_galaxy}
\end{figure}

Figure \ref{volume_galaxy} shows 
 the absolute $r-$band magnitudes of individual compact group member galaxies 
 as a function of redshift.
The sample galaxies are distributed over 
 a redshift range $0.015 < z < 0.212$ and 
 a magnitude range $-22.5 < M_{r} < -16.0$. 
The plot shows no significant difference in redshift and magnitude distribution
 for $N\ge4$ and $N=3$ compact group galaxies.
For comparison, 
 we plot the absolute $r-$band magnitudes of the Hickson compact group members. 
The sample here extends to a higher redshift limit 
 than the Hickson sample. 
 
\begin{figure}
\centering
\includegraphics[width=80mm]{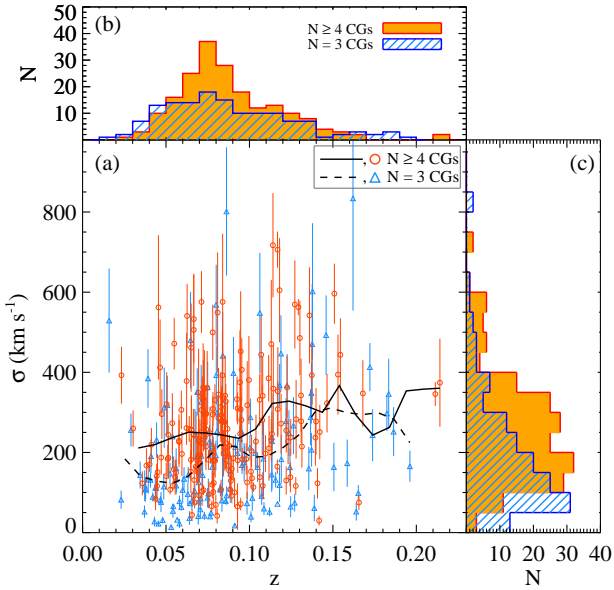}
\caption{
(a) The velocity dispersion ($\sigma$) vs. 
 the redshift for $N\geq4$ compact groups (circles) and for
 $N=3$ compact groups (triangles).
The solid and dashed lines represent 
 the trend after the Nadaraya-Watson kernel regression smoothing 
 for $N\geq4$ and $N=3$ compact groups, respectively. 
(b) and (c) show the redshift and velocity dispersion distributions 
 for $N\geq4$ compact groups (filled histogram) 
 and $N=3$ compact groups (hatched histogram), respectively. }
\label{volume_group}
\end{figure}
 
Figure \ref{volume_group} shows the group velocity dispersion 
 as a function of redshift. 
The median redshift for our sample is $z=0.08$. 
There are more $N=3$ compact groups than $N\ge4$ groups at $z > 0.15$, 
 but the most distant compact groups have  $N\ge4$ at $z=0.211$. 
There is actually no significant difference in the redshift distribution
 between the two types of compact groups. 
The velocity dispersions of $N\ge4$ and $N=3$ compact groups appear
 to increase slightly with redshift,
 but the errors in the velocity dispersion are too large to identify a clear trend. 
 
\begin{figure}
\centering
\includegraphics[width=80mm]{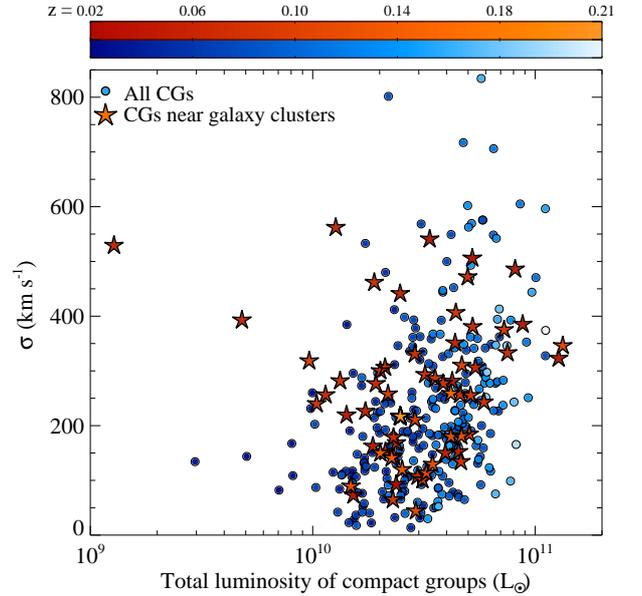}
\caption{Velocity dispersion vs. total r-band luminosity of compact groups.
Circles and starlets show
 compact groups in normal environments and 
 compact groups within rich clusters, respectively. 
Lighter colored symbols represent compact groups at higher redshifts. }
\label{lumsigma}
\end{figure} 
 
To study the cause of the possible slight increase
 in the velocity dispersion of compact groups
 with redshift in Figure \ref{volume_group},
 we plot the velocity dispersion of compact groups 
 as a function of the total group $r-$band luminosity (Figure \ref{lumsigma}).
The total luminosity is the sum of
 $r-$band luminosities of the members.
The velocity dispersion increases with total $r-$band luminosity;
 the correlation tests including 
 Pearson's, Spearman's and Kendall's result in 
 correlation coefficients of 0.27-0.38 
 with the two-sided significance of $\sim0$,
 indicating a weak, but significant correlation. 
The distribution for compact groups near galaxy clusters 
 does not differ from the other groups. 
At higher redshift,
 compact groups containing only low luminosity member galaxies
 are undetectable because the limiting absolute magnitude changes with redshift.
Thus compact groups at higher redshifts tend to have greater total luminosities
 and larger velocity dispersion than nearby compact groups.
The slight increase in the velocity dispersion of compact groups with redshift
 in Figure \ref{volume_group}
 reflects the greater total luminosity of the higher redshift systems.
When we examine the velocity dispersions and 
 total luminosities of the DPOSS II compact groups, 
 the compact groups at higher redshift also have larger velocity dispersion 
 and greater luminosity for the same reason.

\subsubsection{Morphological Content}

We next examine the morphological content of compact groups.  
Both $N\ge4$ and $N=3$ compact groups show 
 larger fractions of early-type galaxies than  late-type galaxies (Table \ref{tab:morph}):  
 $65.3 \pm 1.7\%$ and $56.0 \pm 2.3\%$ of $N\ge4$ and 
 $N=3$ compact group galaxies are early types, respectively. 
In total, $62.1 \pm 1.4\%$ of compact group galaxies are early types. 
This fraction  slightly exceeds  the fraction of early-type galaxies 
 in the Hickson compact groups ($51 \pm 2\%$, \citealp{Hic88}), 
 but it is smaller than the fraction in the DPOSS II compact groups ($81\%$, \citealp{Pom12}). 
However, the early- and late-type galaxies in the DPOSS II compact groups 
 are classified based on H$\alpha$ equivalent width, 
 different from the morphological approach we take.  
Therefore, direct comparison is not possible. 
The fraction of early-type galaxies in our compact groups is 
 similar to the fraction in local galaxy clusters \citep{Par09}. 
 
\begin{table}[t!]
\caption{Morphological Composition of the Sample Compact Groups\label{tab:morph}}
\begin{tabular}{lcc}
\toprule
 & Early-type galaxies & Late-type galaxies \\
\midrule
$N \geq 4$	& 522 ($65 \pm 1.7\%$) & 277 ($35 \pm 1.7\%$) \\
$N = 3$  	    & 235 ($56 \pm 2.3\%$) & 185 ($44 \pm 2.3\%$) \\
Total 		    & 757 ($62 \pm 1.4\%$) & 462 ($38 \pm 1.4\%$) \\
\bottomrule
\end{tabular}
\end{table}

\begin{figure}
\centering
\includegraphics[width=80mm]{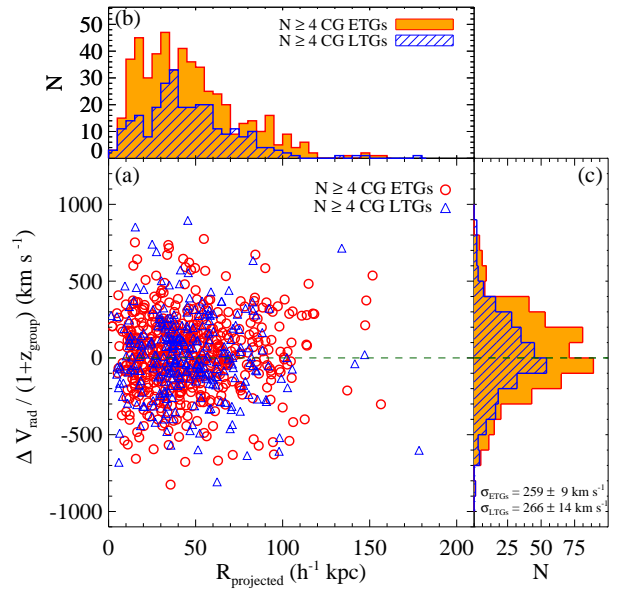}
\caption{
Rest-frame groupcentric radial velocities vs. 
  projected groupcentric distances for $N\geq4$ compact group galaxies. 
Circles and triangles represent early- and late-type galaxies, respectively. 
Their distributions in (b) the projected distances and (c) the radial velocity differences
 are shown with the filled and hatched histogram, respectively. }
\label{rvcg}
\end{figure}

\begin{figure}
\centering
\includegraphics[width=80mm]{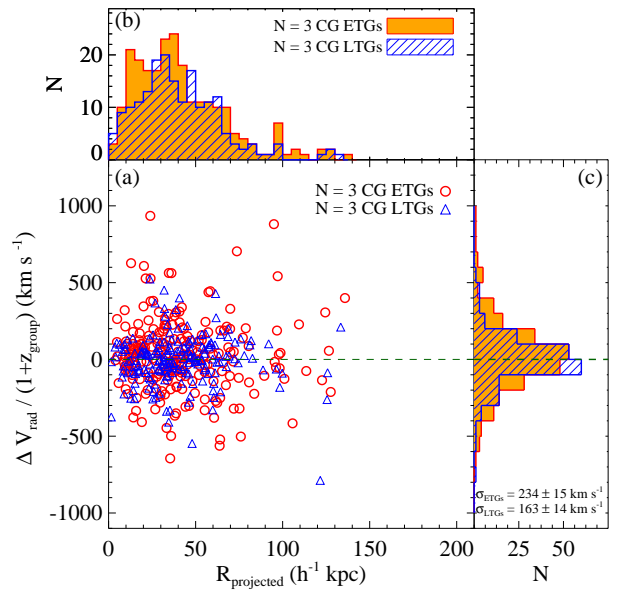}
\caption{
Same as Figure \ref{rvcg}, but for $N=3$ compact groups. } 
\label{rvtr}
\end{figure}

Figure \ref{rvcg} shows 
 the rest-frame groupcentric velocities of galaxies 
 as a function of projected groupcentric radius (i.e. R-v diagram)
 for $N\ge4$ compact groups. 
We use the group centers in Table \ref{tab:grcat} (i.e. R.A., Decl. and redshift) 
 to compute the groupcentric radial velocities and 
 the projected groupcentric distances of member galaxies. 
We then superimpose the groups directly in Figure \ref{rvcg}. 
We distinguish early- and late-type galaxies 
 with different symbols (open circles and triangles).
The distribution of projected groupcentric radius
 for early- and late-type galaxies are similar. 
The Kolmogorov-Smirnov (K-S) test cannot reject the hypothesis that 
 the radial distributions of the two samples are extracted from the same parent population.
The Anderson-Darling (A-D) test 
 gives a result similar to  the K-S test.
The distributions of the rest-frame groupcentric velocities also show
 no significant difference.
The velocity dispersions of early- and late-type galaxies
 for $N\ge4$ compact groups are similar, 
 $259~\pm~9~\kms$ and $266~\pm~14~\kms$, respectively. 
These results differ from galaxy clusters that 
 typically show higher velocity dispersions
 for the more centrally concentrated early-type galaxies relative to late-type galaxies 
 \citep{Col96,Mah99,Hwa08}.
However, \citet{Rin13} showed that 
 the velocity distributions of the blue and red galaxies in galaxy clusters
 are not significantly different, 
 similar to the result we obtain for compact groups.
 
Figure \ref{rvtr} shows a similar R-v diagram for $N=3$ compact groups.
The projected groupcentric distributions for 
 early- and late-type galaxies are similar  
 to the $N\ge4$ compact groups.
However, the velocity dispersion of early-type galaxies is 
 significantly larger than for the late-type galaxies 
 ($234~\pm~15~\kms$ vs. $163~\pm~14~\kms$)
 in contrast with the $N\ge4$ compact groups. 
 
\begin{figure}
\centering
\includegraphics[width=80mm]{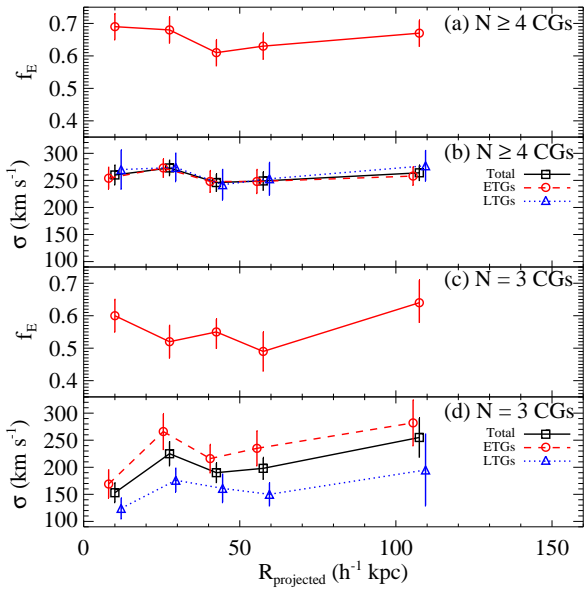}
\caption{
(a) Fraction of early-type galaxies vs. 
 projected distances from the group center for $N\ge4$ compact groups. 
(b) The velocity dispersion of
 all $N\ge4$ member galaxies (square-solid lines), 
 early- (circle-dashed lines), 
 and late-type galaxies (triangle-dotted lines).
 Each point is arbitrarily shifted along the x-axis for clarity. 
(c) and (d) panels are the same as (a) and (b), but for $N=3$ compact groups. } 
\label{rfrac}
\end{figure} 

Figure \ref{rfrac} shows the fraction of early-type galaxies 
 as a function of projected groupcentric distance 
 for $N\ge4$ and $N=3$ compact groups.
We set the bin size to include a similar number of galaxies in each bin.
The fraction of early-type galaxies appears
 to decrease with groupcentric radius in the range
 $0 < R_{projected} < 70 ~h^{-1}$ kpc
 for both $N\ge4$ and $N=3$ compact groups.
However, the fraction in the outermost region 
 at $70 < R_{projected} < 150 ~h^{-1}$ kpc is 
 as high as the fraction in the very inner region.
Because we have a similar number of galaxies in each radial bin,
 this behavior does not simply result from small number statistics.
 
We also examine the velocity dispersion of 
 early- and late-type galaxies
 as a function of groupcentric distance.
The dispersion profiles do not change much
 with groupcentric radius except for the innermost region of $N=3$ compact groups.
This result differs from dispersion profiles 
 for galaxy clusters that typically increase 
 with decreasing clustercentric radius \citep{Mah99, Biv04, Hwa08},
 but on much larger scale.

Figure \ref{rfrac} shows that 
 we can explore the radial dependence of properties of compact groups
 (including the early-type fraction and velocity dispersion)
 in a radial range $0 < R_{projected} < 150 ~h^{-1}$ kpc. 
This small range is interesting because it is hard to sample in any other systems.
The scale is, for example, comparable to or smaller than 
 the typical size of the brightest cluster galaxies ($\sim 100$ kpc, \citealp{New13, Lop14}). 
The morphological properties of galaxies in these small dense regions
 may ultimately provide interesting tests of processes involved in galaxy evolution.
 
\subsubsection{The Abundance of Compact Groups}

\begin{figure*}
\centering
\includegraphics[width=180mm]{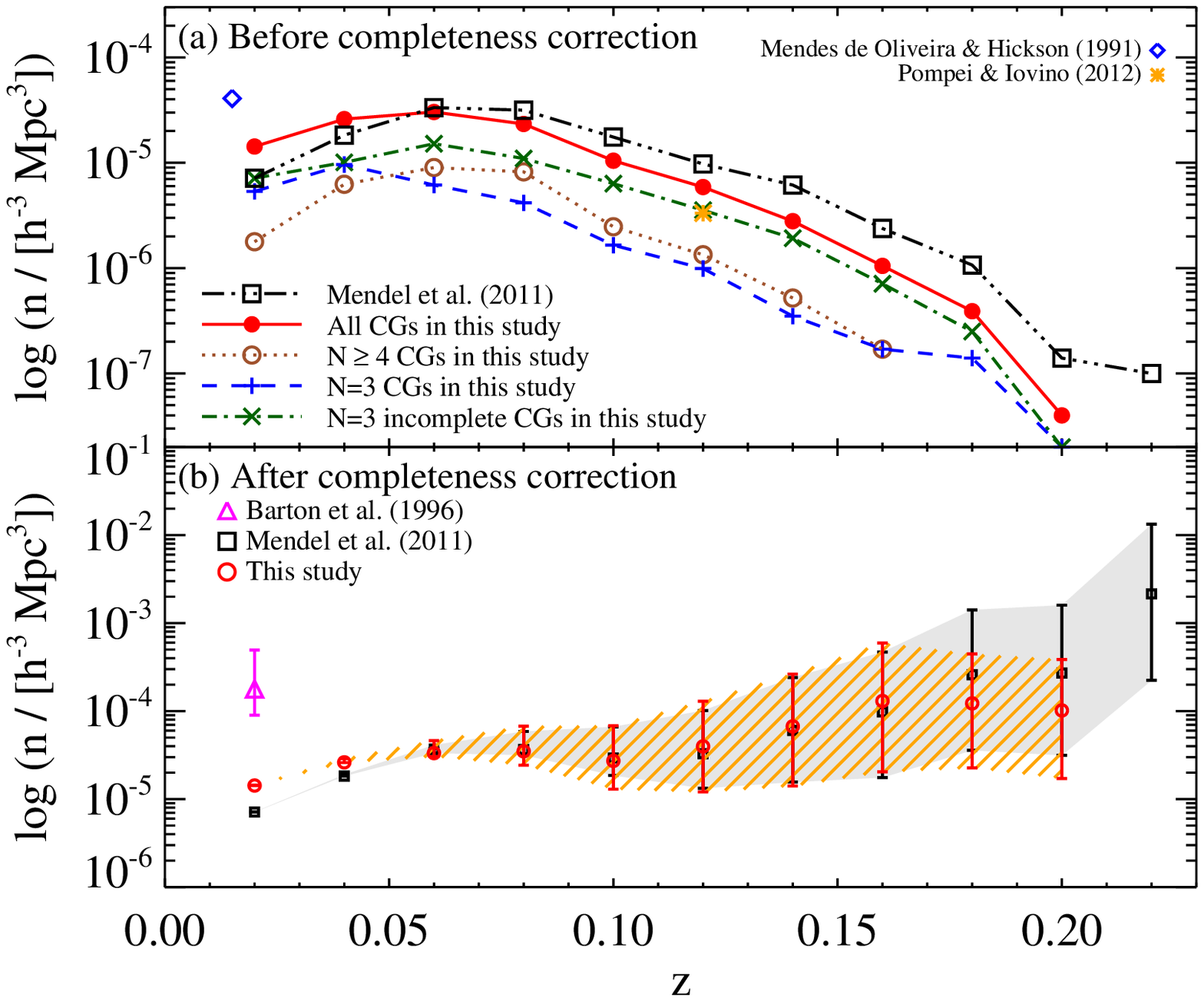}
\caption{
The abundance of compact groups as a function of redshift
 (a) before the redshift effect correction and 
 (b) after the correction. 
The abundances of other compact groups are shown for comparison;
 the Hickson compact groups (diamond, \citealp{MdO91}),
 the CfA2 compact groups (triangle, \citealp{Bar96}), 
 the SDSS DR7 compact groups identified by photometric redshifts (square, \citealp{Men11}), 
 and the DPOSS II compact groups (asterisk, \citealp{Pom12}). 
The symbols are shown at the mean redshift for each group survey. }
\label{abun}
\end{figure*}

Figure \ref{abun} shows
 the abundance of compact groups as a function of redshift.
To compute the abundance,
 we first count all of the compact groups in \citet{McC09}
 including the $N\ge4$ compact groups, $N=3$ compact groups, 
 and $N\geq3$ incomplete compact groups
 in the volume 
 of SDSS DR7 main galaxy survey.
Although the $N\geq3$ incomplete compact groups are not included
 in our compact group catalog,
 they are useful for determining the abundance of $N\ge3$ compact groups.
Because we use a magnitude-limited sample of galaxies to identify compact groups,
 the variation in the absolute magnitude limit as a function of redshift
 affects the observed space density of compact groups.

To correct for this effect,
 we follow the method of \citet{Bar96},
 who computed the abundance of compact groups
 identified by applying a friends-of-friends method
 to the magnitude-limited sample of CfA2+SSRS2 redshift survey data 
 \citep{Gel89, Gio85, daC94}. 
\citet{Bar96} assumed  that the galaxies in their sample compact groups
 are randomly drawn from a magnitude distribution $\overline{\Phi}(M)$.
They then calculated $P_{i}$,
 the probability for detecting $i$th brightest member of a compact group
 in the absolute magnitude range $[M, M+dM]$.
The $P_{i}$ is proportional to $P_{(i-1) < M}(M)\overline{\Phi}(M)dM$,
 where $P_{(i-1) < M}(M)$ is the probability that $i-1$ group members are brighter than $M$.
They derived the $P_{(i-1) < M}(M)$ from the Poisson distribution:
\begin{equation}
 P_{(i-1) < M} = \frac{e^{-\lambda_{M}} \lambda_{M}^{(i-1)}}{(i-1)!},
\end{equation}
 where $\lambda_{M}$ is the average number of galaxies in a group brighter than $M$:
\begin{equation}
 \lambda_{M} = \kappa \int_{-\infty}^{M} \overline{\Phi}(M')dM'.
\end{equation}
Here, $\kappa$ is a normalization parameter, and is a function of redshift.
Finally, they expressed the probability of detecting $\geq i$ group member galaxies:
\begin{equation}
\begin{split}
 P_{detection} (z) &= \frac{1}{A} \int_{-\infty}^{M_{lim(z)}} P_{i}(M)dM \\
                   &= \frac{1}{A} \int_{-\infty}^{M_{lim(z)}} \frac{e^{-\lambda_{M}} \lambda_{M}^{(i-1)}}{(i-1)!} \overline{\Phi}(M)dM,
\end{split}
\end{equation}
where $M_{lim(z)}$ is the limiting absolute magnitudes at the redshift, and
 $A$ is the normalization factor which makes $P_{detection} (z=0) = 1$.

Following the method in \citet{Bar96},
 we construct the selection function for our sample compact groups.
We use the luminosity function determined from nearby galaxies in the SDSS data:
 a Schechter function with
 $\alpha = -0.918 \pm 0.027$ and $M_{star} = -20.31 \pm 0.04$
 from a volume-limited sample of galaxies
 with $0.025 < z < 0.044$ and $M_{r} < -18.0$ \citep{Choi07}.
We use fixed $\kappa$, the median of the measured $\kappa$ for each compact group.
To estimate the uncertainty in the selection function,
 we determine the upper and lower limits for the selection function
 using the 1st and 3rd quartiles of $\kappa$.
Then, we derive the detection probability for compact groups $P_{i}$ using equation (3).

Finally, we calculate the volume number density of compact groups ($n_{cg}$)
\begin{equation}
 n_{cg} = \frac{N}{\frac{\Omega_{survey}}{\Omega_{all~sky}} \int_{z_{i}}^{z_{f}} \frac{4}{3} \pi D_{c}(z)^{3} P_{detection} (z) dz},
\end{equation}
 where $N$ is the total number of $N\ge4$ and $N=3$ compact groups and 
 $N\geq3$ incomplete groups in the redshift range. In equation (5), 
 $\Omega_{survey}$ and $\Omega_{all~sky}$ are, respectively, 
 the solid angles of the SDSS DR7 (8032 deg$^2$) and the full sky,
 and $D_{c}$ is the comoving distance.
We also calculate the 1st and 3rd quartile of the abundance of compact groups 
 using the corresponding $\kappa$.
We take these values as upper and lower limits.
 
Before we apply the completeness correction, 
 we estimate the mean abundance for compact groups 
 at the $z=0.1$ to compare with result from the literature \citep{Men11, Pom12}. 
We use the equation (1) of \citet{Lee04} to estimate the mean abundance. 
The mean abundance for our sample compact groups is 
 $1.76 \times 10^{-5} ~h^{3} ~{\rm Mpc^{-3}}$,
 similar to that for the sample of \citet{Men11}
 ($\sim 2.59 \times 10^{-5} ~h^{3} ~{\rm Mpc}^{-3}$),
 but larger than for the DPOSS II compact groups 
 ($\sim 3.32 \times 10^{-6} ~h^{3}~{\rm Mpc^{-3}}$, \citealp{Pom12}).
We note that 
 the DPOSS II compact group abundance accounts only for `isolated' groups.
The estimate of \citet{Men11} is exceeding our result even though 
 they used only $N\geq4$ compact groups. 
Because \citet{Men11} identified their sample groups using photometric redshifts and
 did not apply a completeness correction, 
 they may overestimate the compact group abundance. 
In addition, 
 several studies have estimated the mean group abundance 
 \citep{MdO91, Iov03, Lee04, deC05, Dia12}, 
 but it is difficult to compare these abundances with our result 
 because of differences in the selection criteria and 
 the lack of a completeness correction. 

The bottom panel shows that 
 the abundance of our sample compact groups 
 as a function of redshift  
 after the completeness correction.
We also plot the abundance for the compact group sample of \citet{Men11}
 after a similar completeness correction. 
We note that the sample from \citet{Men11} is derived from photometric redshifts,
 and only contains $N\geq4$ compact groups. 

The abundances of compact groups in this study and 
 in \citet{Men11} at low redshift (i.e. $z \sim 0.02$) 
 appear smaller than the abundance of the CfA2 compact groups \citep{Bar96}. 
This difference could result from the Hickson's isolation criterion. 
The isolation criterion requires that 
 there be no other galaxies within three magnitudes of the brightest group galaxy
 within the isolation annulus ($R_{G} < R_{GCD} < 3 R_{G}$, 
 where $R_{GCD}$ is groupcentric distance).
This criterion was introduced to avoid very dense regions like cluster cores,
 but it tends to reject some nearby groups.
In general, 
 the spatial extent of nearby groups are larger than for distant groups.  
Thus, 
 the isolation annulus for nearby groups is larger than that for high-redshift groups
 and some nearby compact groups with large spatial extent
 may not be selected as compact groups
 because there are many interlopers within the isolation annulus. 
The compact groups in this study and in \citet{Men11} 
 are identified with Hickson's criterion,
 but the compact groups in \citet{Bar96} are selected with a friends-of-friends algorithm.
We plan to examine this issue further using a method similar to \citet{Bar96},
 but with a large spectroscopic sample of galaxies.   
 
The abundance of our sample compact groups changes little 
 as a function of redshift. 
The sample of \citet{Men11} also changes little with redshift. 
These results are consistent with \citet{Bar96},
 who showed that the abundance of compact groups
 does not change in the redshift range $0.00 < z < 0.03$
 when the compact groups are identified directly from
 a spectroscopic sample of galaxies using the friends-of-friends algorithm.
\cite{Kro15} suggested that
 there would be more compact groups 1 Gyr ago (e.g. $z\sim0.1$)
 than in the current universe
 if compact groups collapse into a single elliptical galaxies
 on a short timescale.
Our results suggest that the space density of compact groups does not
 change significantly with redshift.
Thus compact groups either survive longer than 1 Gyr or
 they are replenished with galaxies from the surrounding region.
 
\subsection{Comparison with Other Compact Group Samples}
To compare the physical properties of the compact groups 
 in this study with those in previous studies,
 we use samples of compact groups 
 based on similar selection criteria and on redshift survey data:
 Hickson compact groups \citep{Hic92} and DPOSS II compact groups \citep{Pom12}. 
The Hickson compact groups are in the redshift range $0.003 < z < 0.333$ and 
 they include member galaxies with $r < 19.5$.
The DPOSS II compact groups span 
 a redshift range $0.044 < z < 0.233$ and
 the magnitudes of the member galaxies are $r < 19.0$.

Both samples are based on Hickson¡¯s selection criteria, 
 but the selection criteria for the DPOSS II compact groups differ slightly from Hickson's.
\citet{Iov03} and \citet{deC05} used: 
 $N(\Delta m <2) \geq 4$, 
 $R_{N} \geq 3 R_{G}$, and 
 $\mu_{G} \leq 24.0 {\rm ~mag ~arcsec}^{-2}$.  
These criteria differ from Hickson's  in several ways. 
First, Hickson used $N(\Delta m < 3)$ rather than $N(\Delta m <2)$, 
 where $N(\Delta m <2)$ refers to the total number of galaxies 
 within 2 mag of the brightest member galaxy. 
Second, $R_{N}$ in Hickson's criteria 
 is the angular size of the smallest circle encompassing no additional galaxies 
 within 3 mag of the brightest group member, 
 while $R_{N}$ for the DPOSS II compact groups is 
 the angular size of the largest circle that includes no additional galaxies 
 within 0.5 mag of the faintest group member. 
The definitions of $R_{G}$ are the same.
Third, the mean surface brightness limit for Hickson's criteria is 
 $26.0 {\rm ~mag ~arcsec}^{-2}$ 
 rather than $24.0 {\rm ~mag ~ arcsec}^{-2}$. 
These differences 
 between the DPOSS II and the Hickson account for
 some of the differences in the physical properties of compact groups in the two samples.

\subsubsection{Physical Properties}

\begin{figure}
\centering
\includegraphics[width=80mm]{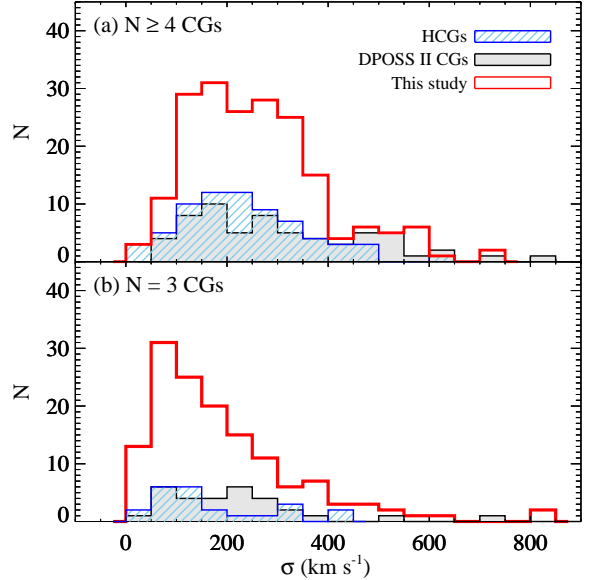}
\caption{Distribution of group velocity dispersion ($\sigma$) for 
 $N\geq4$ compact groups and $N=3$ compact groups in
 our samples (open histogram) compared with 
 the Hickson compact groups (hatched histogram) and 
 the DPOSS II compact groups (filled histogram). }
\label{veldisp}
\end{figure}

We estimate the velocity dispersion for our sample compact groups
 following \citet{Dan80}; 
 the dispersions range from $13 ~\kms$ to $834 ~\kms$ 
 with a typical error of $44 ~\kms$ (Figure \ref{veldisp}).
The median velocity dispersion of all, 
 $N\geq4$, and $N=3$ compact groups are, respectively, 
 $207~\pm~12 ~\kms$, $244~\pm~11~\kms$ and $160~\pm~14 ~\kms$. 
These results are similar to the Hickson compact groups 
 (median($\sigma) = 204~\pm~13~\kms$), 
 but smaller than for the DPOSS II compact groups 
 (median($\sigma) = 251~\pm~22~\kms$). 
When we compare the velocity dispersions of 
 $N\geq4$ and $N=3$ compact groups separately, 
 the velocity dispersions of the DPOSS II compact groups still exceed 
 those for other two samples. 
 
\begin{figure}
\centering
\includegraphics[width=80mm]{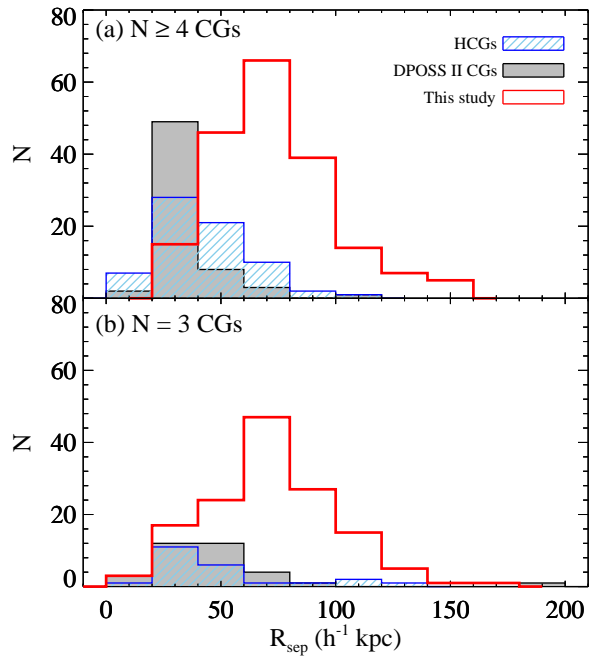}
\caption{Distribution of the median projected separation ($R_{\rm sep}$)
 of member galaxies for 
 (a) $N \geq 4$ compact groups and 
 (b) $N = 3$ compact groups. 
The histograms are as in Figure \ref{veldisp}. }
\label{size}
\end{figure} 
 
Figure \ref{size} shows the 
 distribution of the median projected separation of member galaxies in our sample
 compared with the Hickson and the DPOSS II groups. 
Our sample compact groups have a median projected separation ($R_{\rm sep}$) ranging from 
 11 $h^{-1}$ kpc to 167 $h^{-1}$ kpc;
 this range is similar to that of other compact groups:
 2 $h^{-1}$ kpc to 135 $h^{-1}$ kpc for the Hickson compact groups and 
 12 $h^{-1}$ kpc to 188 $h^{-1}$ kpc for the DPOSS II compact groups.
However, the median projected separation of our sample compact groups
 is larger (median $R_{\rm sep} \sim 72 ~h^{-1}$ kpc) than 
 that for the Hickson compact groups (median $R_{\rm sep}  \sim 39 ~h^{-1}$ kpc) and
 the DPOSS II compact groups (median $R_{\rm sep} \sim 34 ~h^{-1}$ kpc). 
When we compare $N\ge4$ and $N=3$ compact groups separately,
 these differences remain. 
The group radius of our sample compact groups 
 (median $R_{G} \sim 57.8 \pm 1.5 ~h^{-1}$ kpc) is also larger than
 that of the Hickson compact groups (median $R_{G} \sim 38.6 \pm 7.1 ~h^{-1}$ kpc),
 but similar to that of the our parent sample 
 (median $R_{G} \sim 62 ~h^{-1}$ kpc \citealp{McC09}).

\begin{figure}
\centering
\includegraphics[width=80mm]{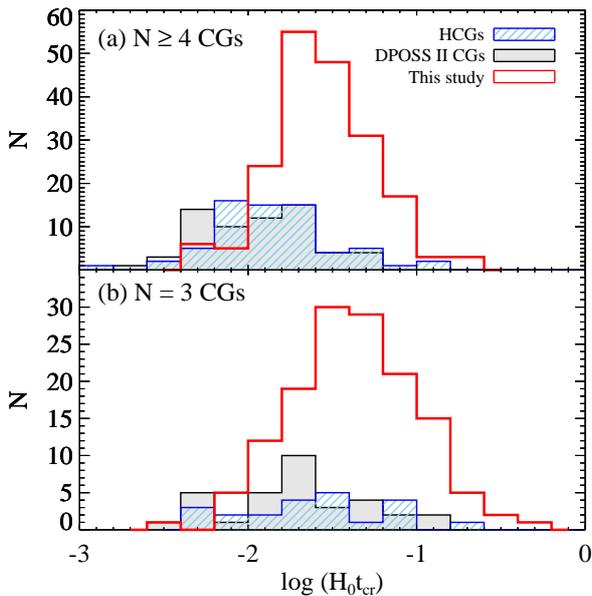}
\caption{Distribution of the crossing time for
(a) $N\geq4$ compact groups and 
(b) $N=3$ compact groups.
The histograms are as in Figure \ref{veldisp}. }
\label{cross}
\end{figure}
 
We also derive the crossing times of the compact groups (Figure \ref{cross}). 
The crossing time is  
\begin{equation}
t_{cr} = \frac{4 R_{\rm sep}}{\pi \sigma_{\rm 3D}},
\end{equation}
 where $R_{\rm sep}$ is the median galaxy-galaxy separation and
 $\sigma_{\rm 3D}$ is the three-dimensional velocity dispersion 
 (see equation (1) and (2) in \citealp{Hic92}). 
The dimensionless crossing time ($H_{0} t_{cr}$) for our sample compact groups
 ranges from 0.004 to 0.469, 
 similar to the distributions of the Hickson and the DPOSS II compact groups. 
The median crossing time ($0.033 \pm 0.003$) for our sample compact groups is 
 larger than for the Hickson ($0.016 \pm 0.131$) and 
 for the DPOSS II ($0.015 \pm 0.002$) compact groups. 
The larger crossing time of our sample compact groups results from
 the larger inter-galaxy separations 
 compared with other compact groups. 
 
\begin{figure}
\centering
\includegraphics[width=80mm]{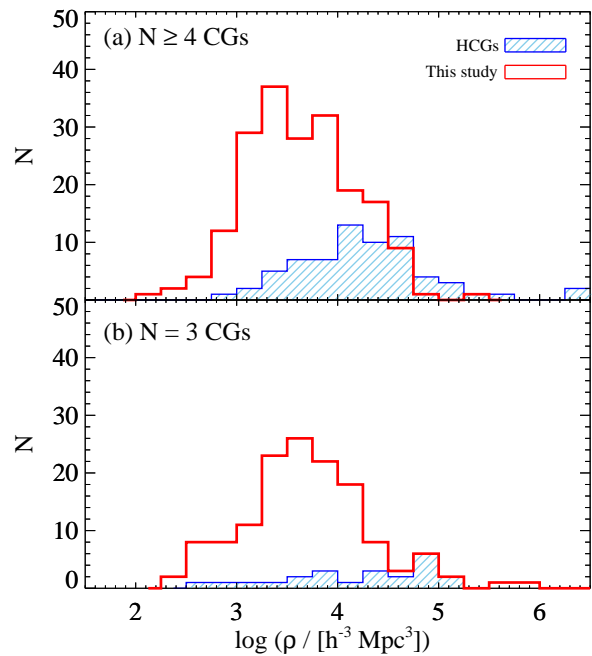}
\caption{Distribution of galaxy number density within groups for 
 (a) $N\geq4$ compact groups and 
 (b) $N=3$ compact groups. 
The histograms are as in Figure \ref{veldisp}, 
 but the density distribution of the DPOSS II compact group
 is not shown here. }
\label{den}
\end{figure} 
 
We compute the group density as in \citet{Bar96}
 \begin{equation}
 \rho = \frac{3N}{4 \pi R_{G}^{3}}
 \end{equation}  
 where $N$ is the number of member galaxies, 
 and $R_{G}$ is the group radius in $h^{-1}$ Mpc.
Figure \ref{den} displays the distribution of densities for our sample
 and for Hickson's sample. 
The number densities in our sample
 are lower than for the Hickson compact groups
 because the sizes of our compact groups are, 
 on average, larger than for the Hickson compact groups.
The median number density for our sample is 
 $\log (\rho/[h^{-3} ~{\rm Mpc}^{3}]) = 3.65$, 
 substantially exceeding the number density for subclusters and subgroups
 in the A2199 superclusters 
 (the median number density at $R < 0.5~R_{200}$ is 
 $\log (\rho/[h^{-3} ~{\rm Mpc}^{3}]) = 1.97$, \citealp{Lee15}). 
Compact groups are indeed much denser environments than galaxy clusters. 
Table \ref{tab:prop} summarizes the physical properties of the compact groups in this study
 compared with the Hickson and the DPOSS II samples.

\begin{table*}[t!]
\caption{Basic Properties of Compact Groups\label{tab:prop}}
\begin{tabular}{lcccccccc}
\toprule
\multirow{2}{*}{Samples} & \multirow{2}{*}{Types} & \multirow{2}{*}{$N_{gr}$} & 
\multirow{2}{*}{$N_{\rm gal}$} & $R_{group}^{\rm a}$ & 
$\log \rho^{\rm a}$ & $R_{\rm sep}^{\rm a}$ & $\sigma^{\rm a}$ & H$_{0}~t_{cr}^{\rm a}$ \\
 &  &  &  & ($h^{-1}$ kpc) & (${h^{3} \rm Mpc}^{-3}$) & ($h^{-1}$ kpc) & ($\kms$) &  \\
\midrule
This study                    & Total           & 332 & 1129 & $57.8 \pm 1.5$ & $3.65 \pm 0.03$ & $71.7 \pm 1.4$ & $207. \pm 12.$ & $0.033 \pm 0.003$\\
($0.016 < z < 0.211$) & $N \geq 4$ & 192 &  799 & $62.4 \pm  2.0$  & $3.64 \pm 0.04$ & $70.5 \pm 1.9$ & $244. \pm 11.$ & $0.028 \pm 0.002$ \\
			                        & $N = 3$ 	     & 140 &  420 & $54.1 \pm  2.4$  & $3.66 \pm 0.05$ & $73.4 \pm 2.4$ & $160. \pm 14.$ & $0.042 \pm 0.005$ \\
Hickson Compact Groups & Total        & 92  &  382 & $38.6 \pm  7.1$  & $4.27 \pm 0.09$ & $39.8 \pm 2.4$ & $204. \pm 13.$ & $0.016 \pm 0.131$ \\
  ($0.003 < z < 0.333$) & $N \geq 4$ & 69 &  313 & $40.1 \pm  9.1$  & $4.21 \pm 0.11$ & $39.9 \pm 2.6$ & $209. \pm 15.$ & $0.012 \pm 0.138$ \\
                                      & $N = 3$ 	  & 23 &   69   & $30.2 \pm 5.6$   & $4.4 \pm 0.17$ & $38.0 \pm 6.7$ & $123. \pm 26.$ & $0.025 \pm 0.325$ \\
DPOSS II Compact Groups & Total 	        &  96 &  370 &  $--$                    & $--$                 & $34.2 \pm 2.3$ & $251. \pm 22.$ & $0.015 \pm 0.002$ \\
  ($0.044 < z < 0.233$)   & $N \geq 4$ &  63 &  271 & $--$                    & $--$                 & $31.4 \pm 1.9$ & $278. \pm 32.$ & $0.012 \pm 0.002$ \\
                                        & $N = 3$ 	 &  33 &   99  & $--$                   & $--$                  & $40.4 \pm 5.4$ & $215. \pm 30.$ & $0.020 \pm 0.006$ \\
\bottomrule
\end{tabular}
\tabnote{$^{\rm a}$ The median of each parameter are listed. The errors are $1\sigma$ standard deviations derived from 1000 bootstrapping resamplings.}
\end{table*}

\subsubsection{Local environments of compact groups}

\begin{figure}
\centering
\includegraphics[width=80mm]{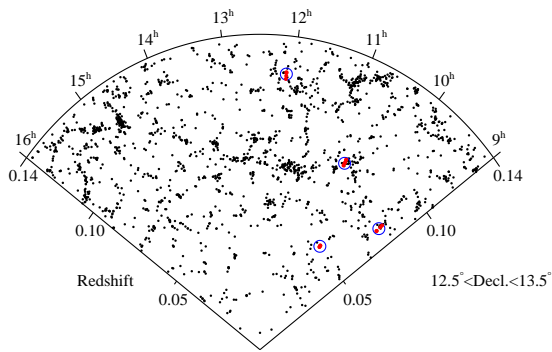}
\caption{
Example cone diagram for a slice of
 $9^{h} < \alpha_{2000} < 16^{h}, ~12.5^{\circ} < \delta_{2000} < 13.5^{\circ}$,
 and $0.00 < z <0.14$.
Large open and small filled circles indicate
 compact groups and their member galaxies, respectively.
Small dots denote SDSS galaxies in a volume limited sample
 with $M_{r} < -20.5$ and $0.01 < z < 0.14$.
The $\Sigma_{5}$ for the compact groups from the right are 
 $\Sigma_{5} = 16.08, 0.07, 0.67$ and 0.19. }
\label{cone}
\end{figure}

\begin{figure}
\centering
\includegraphics[width=80mm]{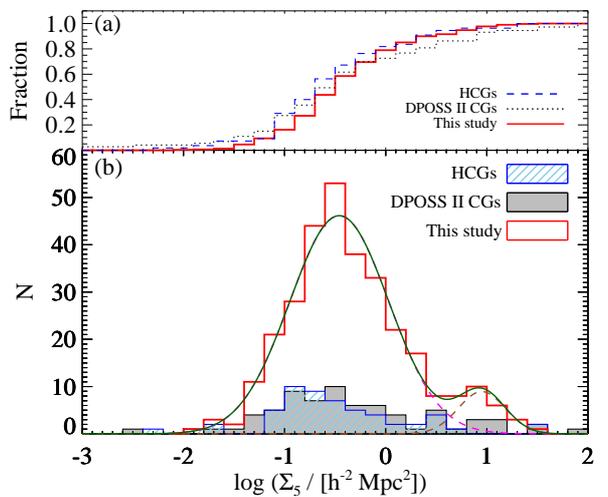}
\caption{
(a) Cumulative distribution of the surrounding surface number density 
 ($\Sigma_{5}$) and (b) the distribution of $\Sigma_{5}$ for
 our sample compact groups (solid line, open histogram) 
 the Hickson compact groups (dashed line, hatched histogram),
 the DPOSS II compact groups (dotted line, filled histogram). } 
\label{env}
\end{figure}

Figure \ref{cone} displays 
 an example of the spatial distribution of compact groups
 relative to the surrounding large-scale structure
 for a slice of $9^{h} < \alpha_{2000} < 16^{h}$ and
 $12.5^{\circ} < \delta_{2000} < 13.5^{\circ}$.
We choose this slice 
 to show various environments of compact groups 
 even though there are only four compact groups in this thin slice. 
To show homogeneous structures of galaxies regardless of redshift,
 we plot galaxies in a volume-limited sample
 with $M_{r} < -20.5$ and $0.00 < z < 0.14$
 (see the large box in Figure \ref{volume_galaxy}).
As expected based on previous studies of smaller samples 
 \citep{Roo94, Ram94, Rib98, And05, Pom12}, 
 the environments of compact groups are diverse (Figure \ref{env}).

In Table \ref{tab:grcat}, 
 we examine the number of compact groups located near galaxy clusters. 
Using NED and requiring 
 $|v_{group} - v_{cluster}| < 3000~\kms$ and $R_{projected} < 1~h^{-1}$ Mpc,
 only 69 (21\%) of our sample compact groups are near known massive clusters.
With more relaxed criteria, 
 $|v_{group} - v_{cluster}| < 6000~\kms$ and 
 $R_{projected} < 1~h^{-1}$ Mpc \citep{Men11}, 
 the number of compact groups near known massive clusters changes little to 80 (i.e. 24\%).
These fractions are smaller than those
 in previous studies; 
 e.g., 35\% in \citet{Pom12} and 50\% in \citet{Men11} based on similar criteria.
\citet{Men11} used  compact groups identified with photometric redshifts 
 and $N > 4$ galaxy groups from the SDSS DR7 \citep{Tag10}
 to study the local environments of compact groups.
In contrast, we use only compact groups with complete spectroscopic redshifts
 and known massive galaxy clusters listed in NED; 
 thus our criteria are stricter than in other studies. 
The more restrictive criteria result in 
 the lower fraction of compact groups within known massive clusters. 

We also investigate the local environments of our sample compact groups 
 using the parameter, $\Sigma_{5}$,
 and compare them with the environments of the Hickson and the DPOSS II compact groups. 
$\Sigma_{5}$ is the surface number density defined as 
 $\Sigma_{5} = 5 (\pi D_{p,5}^{2})^{-1}$,
 where $D_{p,5}$ is the projected distance 
 between the center of the compact group and the fifth nearest neighbor galaxy. 
We use galaxies in the volume-limited sample 
 to compute the nearest neighbor densities
 and to make a fair comparison regardless of redshift.
We use neighbor galaxies with relative velocities $\Delta v < 1500~\kms$, 
 and compute $\Sigma_{5}$ relative to the center of each compact group. 
We obtain $\Sigma_{5}$ for 309 groups within the volume-limited sample.

Figure \ref{env} shows the $\Sigma_{5}$ distribution for
 our sample compact groups along with the Hickson and the DPOSS II compact groups 
 for the redshift range $0.01 < z < 0.14$. 
Some groups in our study are 
 in denser environment than the Hickson compact groups. 
Indeed, the A-D test rejects the hypothesis that 
 the distributions of all three samples are extracted from the same parent population
 with a low p-value ($< 0.1$). 

Figure \ref{env} suggests 
 that the $\Sigma_{5}$ distribution of compact groups
 can be divided into two
 as many previous studies suggested:
 `isolated' and `embedded' compact groups. 
To examine the multiplicity of the $\Sigma_{5}$ distribution,
 we use a statistical test, the Gaussian mixture model (GMM, \citealp{Mur10}).
The GMM evaluates whether the data are more consistent
 with a multimodal Gaussian distribution 
 rather than a unimodal Gaussian distribution.
If the data consist of multiple populations,
 the GMM returns
 1) a low parametric bootstrap method probability,
 2) a large separation ($D > 2$) between multiple Gaussian peaks,
 3) a negative {\it kurtosis} of the input distribution, and
 4) a larger enhancement of the likelihood for the multimodal case than for the unimodal case ($-2 ~\ln (L_{\rm unimodal} / L_{\rm multimodal})$).

We assume that there are two populations of compact groups
 with high and low $\Sigma_{5}$, and apply the GMM.
The GMM test indicates that the $\Sigma_{5}$ distribution
 may have a bimodal distribution with
 low probability $p = 6.97 \times 10^{-6}$,
 {\it kurtosis} $k = 0.186 \pm 0.824$,
 large separation between peaks $D = 3.68 \pm 0.53$, and
 large enhancement of the likelihood 
 $-2 ~\ln (L_{\rm unimodal} / L_{\rm multimodal}) = 29.3$.
This means that the $\Sigma_{5}$ distribution is consistent with bimodal distribution.
The two populations of compact groups are divided at
 $\log (\Sigma_{5}) = 0.62$, and
 91\% (281 out of 309) of compact groups belong to a population with low $\Sigma_{5}$.
If we accept this bimodal distribution of $\Sigma_{5}$ and
 $\Sigma_{5}$ traces  the local environments of compact groups,
 only $\sim 9$\% of compact groups are in dense environments.
This fraction is much lower than in previous studies ($33 - 50 \%$),
 but our division is based on a statistical test that is very strict compared with 
 other studies based on different local environment indicators 
 (e.g. distance to nearby galaxy groups).
Hereafter, we refer the compact groups in dense environments as `embedded' groups,
 and the others as `isolated' groups.

\begin{figure*}
\centering
\includegraphics[width=160mm]{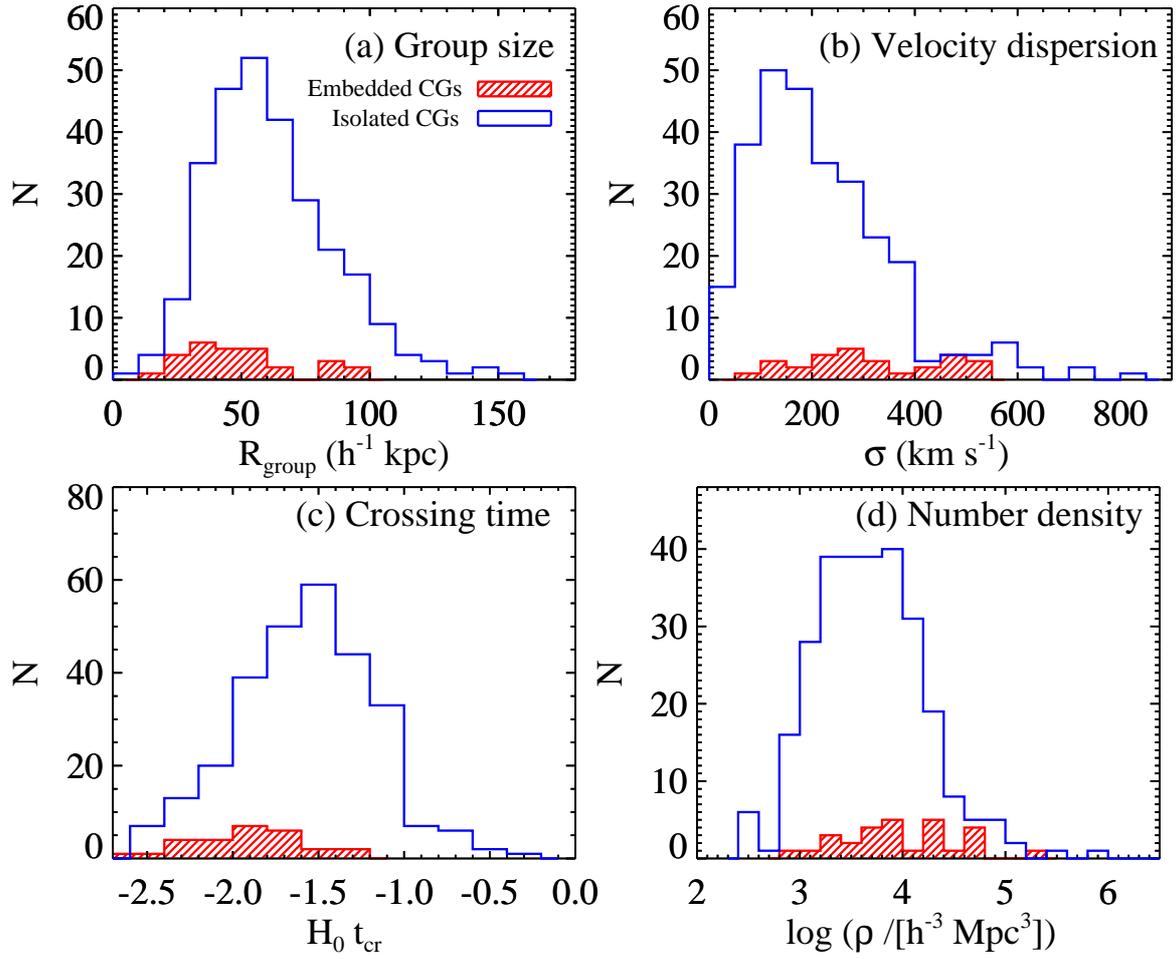}
\caption{The physical properties of embedded (hatched histograms) 
 and isolated (open histograms) compact groups
 including (a) group radius, (b) velocity dispersion, (c) crossing time, and (d) density. 
The embedded groups represent compact groups that have $\log (\Sigma_{5})$ 
 larger than 0.62. }
\label{embsig}
\end{figure*}

Figure \ref{embsig} shows the physical properties of
 the isolated and the embedded compact groups
 including group radius, velocity dispersion, crossing time, and number density.
The median size of embedded groups ($43.3 \pm 4.0 ~h^{-1}$ kpc) is smaller
 than for isolated groups ($57.8 \pm 1.4 ~h^{-1}$ kpc).
The discrepancy also exists even 
 when we compare the $N\ge4$ and $N=3$ compact groups separately.
This result is consistent with previous studies based on different compact group samples \citep{Men11, Dia15}.
The smaller sizes of embedded groups
 result in lower number densities
 than for isolated groups (Figure \ref{embsig} (d)).
The size and density distributions of the two groups are drawn from
 significantly different distributions
 as the A-D test suggests with low p-value $< 0.01$.
The median velocity dispersion of embedded groups ($316 \pm 25~\kms$)
 is significantly larger than that of isolated groups ($219 \pm 8~\kms$).
The distributions of velocity dispersions of the two groups are also different.
This is also consistent with the results in \citet{Pom12}
 for the DPOSS II compact groups.
Therefore,
 the median crossing time of embedded compact groups is
 shorter than the isolated compact groups.
The distributions of crossing time for the two groups
 are also significantly different (p-value from A-D test $<< 0.01$).
When we divide the compact groups
 into isolated and embedded systems based on the distance
 to nearby galaxy clusters,
 the difference in physical properties between the two types remains.

Compact groups based on Hickson's selection criteria,
 including our original catalog of \citet{McC09},
 reflect a selection bias against dense local environments
 as a result of the isolation criterion.
In other words,
 embedded compact groups
 may not fully represent the compact group population in high-density regions
 because many compact groups are missed in or near high-density regions
 as a result of the isolation criterion. 
The impact of this criterion is a function of the
 redshift of the group \citep{Bar96}. 

\section{SUMMARY}

By measuring new redshifts and incorporating redshifts from SDSS DR12 and other literature,
we construct a catalog of 
 192 $N \geq 4$ compact groups with 799 member galaxies and
 140 $N=3$ complete compact groups with 420 member galaxies at $0.01 < z < 0.21$. 
In this catalog,
 all member galaxies have spectroscopic redshifts. 
To date this catalog is the largest spectroscopically complete sample of these unusually dense systems. 
We explore the physical properties of the groups in this catalog and compare them with previous samples. 

We examine the redshift dependence of physical properties of compact groups
 in the redshift range $0.01 < z < 0.21$. 
The velocity dispersion of compact groups changes little with redshift,
 indicating no significant evolution of dynamical masses of compact groups in this redshift range.
The space density of compact groups also shows 
 no significant change with redshift. 
Thus it appears 
that either compact groups can survive longer than 1 Gyr or
 they continually reform by accreting new members from their surroundings. 
  
The early-type fraction in our sample compact groups is 62\%,
 slightly exceeding the fraction in the Hickson compact groups. 
We superimpose all of the compact groups in our sample to investigate
 the radial behavior of the velocity dispersion and morphological fraction.
The velocity dispersion of early- and late-type galaxies are similar 
 in $N\ge4$ compact groups, 
 but the dispersion for the early-type galaxies is larger than 
 for late-type galaxies in $N=3$ compact groups. 
The velocity dispersions of early- and late-type galaxies in compact groups 
 do not change much as a function of groupcentric radius. 
Compact groups enable examination of these issues at a galaxy density and 
 spatial scale that are hard to access with any other systems.
 
We compare the catalog we construct with the Hickson and the DPOSS II samples
 that also have complete spectroscopy.
We compare sizes, number densities, velocity dispersions and environments 
 as measured by the fifth nearest neighbor to the group. 
The physical properties of our sample groups are 
 similar to those for the Hickson compact groups, 
 but they differ from those of the DPOSS II compact groups.
The differences result from differences in the selection criteria for the DPOSS II and
 the Hickson compact groups. 
The parent catalog we use, \citet{McC09}, is based on Hickson's criteria.

The local environments of compact groups are diverse. 
The $\Sigma_{5}$ distribution of compact groups is bimodal
 and 9\% of compact groups are located in the denser region. 
This `embedded' group fraction is lower than previous studies
 based on different local density tracers. 
The embedded compact groups are smaller and have larger velocity dispersion
 than the isolated compact groups on average. 

Compact groups are a fascinating laboratory for studying galaxy evolution. 
Examination of the number density of these systems over a larger redshift range and 
 comparison with simulations may further constrain 
 the formation and evolution of these systems.
It is also important to clarify
 the subtle issues in the identification of the compact systems. 
Further exploration of identification directly 
 from complete spectroscopic surveys in the nearby and 
 moderate redshift universe would provide a further foundation for understanding
 the nature of these systems.

\acknowledgments
We thank the anonymous referee for a very prompt report.
We thank Perry Berlind and Michael Calkins, 
 the remote observers at the Fred Lawrence Whipple Observatory,
 Jessica Mink, who processed the spectroscopic data,
 and all FAST queue observers who took data for this program. 
This paper uses data products produced by the OIR Telescope Data
 Center, supported by the Smithsonian Astrophysical Observatory.
This work was supported by the National Research Foundation of Korea (NRF) grant
 funded by the Korea Government (MSIP) (No.2013R1A2A2A05005120).
J.S. was supported by 
 Global Ph.D. Fellowship Program through an NRF funded by the MEST (No. 2011-0007215). 
The research of M.J.G. is supported by the Smithsonian Institution.
AD acknowledges partial support from the INFN grant InDark, the grant
Progetti di Ateneo TO Call 2012 0011 'Marco Polo' of the University of
Torino and the grant PRIN 2012 ``Fisica Astroparticellare Teorica'' of
the Italian Ministry of University and Research.
G.H.L. acknowledges the support by the National Research
Foundation of Korea (NRF) Grant funded by the Korean
Government (NRF-2012-Fostering Core Leaders of the Future
Basic Science Program).




\end{document}